\documentclass{article}

\newcommand{\liff}{\leftrightarrow}

\newcommand{\ga}{\alpha} \newcommand{\gb}{\beta}
\newcommand{\gc}{\gamma} \newcommand{\gd}{\delta}

\newcommand{\GC}{\Gamma}
\newcommand{\sbs}{\subseteq} 
 \newcommand{\lto}{\rightarrow}

 \newcommand{\cl}{{\cal L}}
 
 \newcommand{\cp}{{\cal P}}

 \newcommand{\cd}{{\cal D}}
\newcommand{\vd}{\vdash}

\newtheorem{theorem}{\bf Theorem}
\newtheorem{lemma}[theorem]{\bf Lemma}
\newtheorem{proposition}[theorem]{\bf Proposition}
\newtheorem{corollary}[theorem]{\bf Corollary}

\newenvironment{definition}{\par\medskip\addtocounter{theorem}{1}%
  \noindent{\bf Definition \arabic{theorem}}\ }{\medskip}

\newenvironment{example}{\par\medskip\addtocounter{theorem}{1}%
 \noindent{\bf Example \arabic{theorem}}\quad}{\medskip}

\newcommand{\qed}{\vrule height5pt width3pt depth0pt}

\newenvironment{proof}{\noindent {\it Proof.}}{{\nobreak\hfill \qed \par
\medbreak}}

\newenvironment{atheorem}{\noindent {\bf Theorem} \em}{\medskip}
\newenvironment{alemma}{\noindent {\bf Lemma} \em}{\medskip}
\newenvironment{aproposition}{\noindent {\bf Proposition} \em}{\medskip}
\newenvironment{acorollary}{\noindent {\bf Corollary} \em}{\medskip}

\newcommand{\ev}{\makebox[1.2em]{$\rule{0.1mm}{2.5mm}\hspace{-1.1mm}\sim$}}
\newcommand{\notev}{\makebox[1.2em]{$\rule{0.1mm}{2.5mm}\hspace{-1.1mm}\not\sim$}}

\title{Ordering-based Representations of Rational Inference\thanks{Work  supported by
Training through Research Contract No. ERBFMBICT950324 between the European
Community and Universit\`{a} degli Studi di Roma ``La Sapienza''.}}

\author{Konstantinos Georgatos\\
Dipartimento di Informatica e Sistemistica\\ Universit\`a di Roma ``La
Sapienza''\\ Via Salaria 113, Roma 00198\\ Italy}

\begin{document}
\maketitle

\begin{abstract}
Rational inference relations were introduced by Lehmann and Magidor as the ideal
systems for drawing conclusions from a conditional base. However, there has been no
simple characterization of these relations, other than its original representation
by preferential models. In this paper, we shall characterize them with a class of
total preorders of formulas by improving and extending G\"ardenfors and Makinson's
results for expectation inference relations.  A second representation is
application-oriented and is obtained  by considering a class of consequence
operators that grade sets of defaults according to our reliance on them. The finitary
fragment of this class of consequence operators has been employed by recent default
logic formalisms based on maxiconsistency.
\medskip

\end{abstract}

\section{Introduction}

A recent breakthrough in nonmonotonic logic is the beginning of study of
nonmonotonic consequence through postulates for abstract nonmonotonic
consequence relations, using Gentzen-like context-sensitive sequents (\cite{GAB85},
\cite{MAK89}, \cite{KLM90}). The outcome of this research turns out to be valuable in
at least two ways

\begin{itemize}

\item it provides a sufficiently general {\em axiomatic\/} framework for comparing and
classifying nonmonotonic formalisms, and

\item it gave rise to new, simpler, and better behaved systems for nonmonotonic
reasoning, such as cumulative (\cite{GAB85}), preferential (\cite{KLM90}), and rational
(\cite{LM92}) inference relations.

\end{itemize}

It is unfortunate that these new inference relations  enjoy
only one, semantical, representation; that of preferential
models (\cite{SHO87}). We have that {\em preferential\/}, {\em preferential
transitive\/}, and {\em preferential modular\/} or {\em ranked\/} models characterize
cumulative, preferential and rational inference relations, respectively
(\cite{KLM90}, \cite{LM92}). An additional second-order constraint must be imposed on
these models, called {\em stoppering\/} or {\em smoothness\/}.  However, this
modeling is insufficient because in order to employ the above inference relations, one
must be able to generate them. This is crucial when we want to design a system that
reasons  using the above inference relations. In such a case, one comes up with a set
of rules or defaults that one wants to apply, imposes a prioritization on them, and
provides a mechanism which ensures that answers are derived according to these
inference relations. This is exactly the proof-theoretic approach expressed by default
logic. However, no similar proof-theoretic notion is provided in the above framework.

In this paper, we  offer two new, alternative  representations
 for rational inference. The first  representation is  algebraic
and  obtained through a simple class of orderings of formulas, called {\em
rational\/} orderings. The second  representation is proof-theoretic and
obtained through a class of consequence operators based on the way we handle defaults,
called {\em ranked consequence operators\/}. Moreover, a correspondence result
between these classes is established.

The first link between nonmonotonic inference relations and a class of orderings of
formulas was given by G\"ardenfors and Makinson in~\cite{GM94}. However, the
nonmonotonic system defined by an ordering of formulas is not one of the previously
mentioned systems, but a translation of the well-known belief revision AGM axioms
(\cite{AGM85}) into nonmonotonic reasoning, called {\em expectation\/}  inference
relations. Expectation inference relations are rational inference relations together
with a rule called {\em Consistency Preservation\/}. Moreover, G\"ardenfors and
Makinson's representation of expectation inference relations  with orderings of
formulas is not appropriate, in the sense that the correspodence is not bijective.
(Two orderings of formulas can generate the same inference relation.) So,  two
questions remain open. Namely,

\begin{itemize}
\item is there a way to generate one of the independently motivated
nonmonotonic inference relations (cumulative, preferential, rational) with a class
of orderings of formulas?, and
\item can the correspondence be bijective?
\end{itemize}

We  answer affirmatively both questions for rational inference. Our approach
is the following. We  study the rule of Consistency Preservation and, by
giving it a precise syntactic characterization, show that its role is
insignificant in the context of preferential reasoning. Drawing from this
intuition, we  introduce  new defining conditions relating the classes of
orderings of formulas and nonmonotonic consequence relations and show that
G\"ardenfors-Makinson orderings are in bijective correspondence with
rational inference. Moreover, we  introduce a smaller class of orderings,
which, under our translation, is in bijective correspondence with the
G\"ardenfors-Makinson expectation inference relations. This is how the first
representation result for rational inference is obtained. This result adds
to a long tradition of defining nonmonotonic logics with orderings of
formulas (\cite{DUP91}, \cite{PEA88}, \cite{GM94}, \cite{ROT91}).

The above representation result is more ``constructive'' than the semantical
completeness of preferential models. However, rational orderings must have a concise,
constructive representation. To this end, we  encode a  natural  way of applying
defaults into a new class of consequence relations, called  {\em ranked consequence
operators\/}. Each member of this class generates a rational  ordering, and
conversely, hence  the class of ranked consequence relations coincides with that of
rational inference. Also, we show how previous default logic systems  in the literature
(\cite{NEB91}, \cite{NEB94}) reduce to our framework.

The above results pave the way towards a study of nonmonotonicity through orderings
of formulas, allow us to translate previous work in belief revision into the context
of nonmonotonic reasoning, and provide a framework for designing default systems
obeying rational inference.

The plan of this paper as follows. In Section~\ref{sec:preliminaries}, we briefly
introduce the relations under study, explain the rule of Consistency Preservation, and
provide a characterization for this rule. In Section~\ref{sec:orderings}, we
introduce the orderings, their translations and our first representation theorem. In
Section~\ref{sec:ranked}, we  introduce ranked consequence operators and  our
second representation theorem. In Section~\ref{sec:rational-default}, we show how one
can generate a ranked consequence operator given a prioritized family of sets of
defaults and, in Section~\ref{sec:conclusion}, conclude. A preliminary version of the
first half of this paper appeared in~\cite{KG95a}. Results from the second half were
announced in~\cite{KG95c}.

\section{Shifting underlying entailment}
\label{sec:preliminaries}

Before going to the main result of this section, we shall make a brief
introduction to the nonmonotonic consequence relations under study.
Assume a language $\cl$ of propositional constants closed under the boolean
connectives $\lor$ (disjunction), $\land$ (conjunction), $\neg$ (negation) and
$\to$ (implication). We shall use greek letters  $\ga$, $\gb$, $\gc$, etc.
for propositional variables. We shall also use $\ga\ev\gb$ , read as ``$\ga$
normally entails $\gb$'', to denote the nonmonotonic consequence relation
($\ev\sbs\cl\times\cl$).  Before we present the first set of rules for $\ev$, we need a
symbol for a classical-like entailment. We shall use $\vd$. The relation $\vd$ need
not be that of classical propositional logic. We require that $\vd$ includes
classical propositional logic, satisfies compactness (i.e., if $X\vd\gb$ then
there exists a finite subset $Y$ of $X$ such that $Y\vd\gb$)\footnote{We write
$X,\ga\vd\gb$ for $X\cup\{\ga\}\vd\gb$.}, the deduction theorem (i.e., $X,\ga\vd\gb$ if
and only if $X\vd\ga\to\gb$) and disjunction in premises (i.e., if $X,\ga\vd\gb$ and
$X,\gc\vd\gb$ then $X,\ga\lor\gc\vd\gb$). The reader will notice that these
are the only properties we make use of in the subsequent proofs. We shall denote
the consequences of $\ga$ with $Cn(\ga)$ and $C(\ga)$ under $\vd$ and $\ev$,
respectively.

The rules mentioned in the following are presented in Table~\ref{table:rules}.
For a motivation of these rules, see \cite{KLM90} and \cite{MAK94}. (The latter serves
as an excellent introduction to nonmonotonic consequence relations.)

\begin{table}[t]
\caption{\label{table:rules} Rules for Preferential, Rational and
Expectation Inference}
\vspace{2pt}
\begin{center}
\begin{minipage}{4.5in}
{\small
\begin{tabular}{cl}

$\displaystyle \frac{\ga \vd \gb}{\ga\ev\gb}$ &
{\footnotesize (Supraclassicality)}
\vspace{.1in} \\

$\displaystyle \frac{\ga\vd\gb \qquad \gb\vd\ga \qquad \ga\ev\gc}{\gb\ev\gc}$ &
{\footnotesize (Left Logical Equivalence)} \vspace{.1in}\\

$\displaystyle \frac{\ga\ev\gb \qquad \vd\gb\to\gc}{\ga\ev\gc}$ &
{\footnotesize(Right Weakening)} \vspace{.1in}\\

$\displaystyle \frac{\ga\ev\gb \qquad \ga\ev\gc}{\ga\ev\gb\land\gc}$
& {\footnotesize(And)} \vspace{.1in}\\

$\displaystyle \frac{\ga\ev\gb \qquad
\ga\land\gb\ev\gc}{\ga\ev\gc}$ & {\footnotesize(Cut)} \vspace{.1in}\\

$\displaystyle \frac{\ga\ev\gb \qquad
\ga\ev\gc}{\ga\land\gb\ev\gc}$ & {\footnotesize(Cautious
Monotonicity)}\vspace{.1in}\\

$\displaystyle \frac{\ga\ev\gc \qquad
\gb\ev\gc}{\ga\lor\gb\ev\gc}$ & {\footnotesize(Or)} \vspace{.1in}\\

$\displaystyle \frac{\ga\notev\neg\gb \qquad
\ga\ev\gc}{\ga\land\gb\ev\gc}$ & {\footnotesize(Rational
Monotonicity)}\vspace{.1in}\\

$\displaystyle \frac{\ga \ev \bot}{\ga\vd\bot}$ &
{\footnotesize (Consistency Preservation)}

\end{tabular}
}
\end{minipage}

\end{center}
\end{table}

\begin{definition}
Following (\cite{KLM90}, \cite{LM92}, \cite{GM94}), we shall say that a relation $\ev$
on $\cl$ is an {\em inference relation (based on $\vd$)\/} if it satisfies
Supraclassicality, Left Logical Equivalence, Right Weakening, and And. We shall call
an inference relation $\ev$ {\em preferential\/} if
it satisfies, in addition, Cut, Cautious Monotonicity, and Or. We shall call
an inference relation $\ev$ {\em rational\/} if
it is preferential and satisfies, in addition,  Rational Monotonicity. Finally, we
shall say that
$\ev$ is an {\em expectation\/} inference relation (based on $\vd$) if it is a
rational  and satisfies, in addition, Consistency Preservation.

\end{definition}

The most controversial of these rules is Rational Monotonicity, which, moreover, is
non-Horn. For a plausible counterexample, see~\cite{STA94}.

 Expectation inference relations correspond to the so-called AGM postulates for belief
revision~(\cite{AGM85}), as it was shown in~\cite{MG91}, and only differ from rational
relations in that they satisfy the following rule, called {\em Consistency
Preservation}:
$$\frac{\ga\ev\bot}{\ga\vd\bot}$$
where $\vd$ is classical entailment. Consistency Preservation says that a logically
not false belief cannot render our set of beliefs inconsistent. This makes a
difference between the two classes, in the following sense. Using rational inference, I
can
 rely on an inference such as $\gb\ev\bot$, where $\gb$ is the statement ``I
am the Queen of England''. On the other hand, expectation
inference would not allow that, since, even if I am certain I am not the Queen of
England, one could think a world where I could have been. This becomes more important
if, instead of a belief set, one considers a conditional base. For example, consider a
database for air-traffic. The statement ``two airplanes are scheduled to arrive at the
same time and land on the same place'' should infer inconsistency on this database,
although it is not a falsity. More examples can be drawn from physical laws.  This
means that rational inference is the logic of ``hard constraints'', that is of
statements (not necessarily tautologies) I cannot revise without deconstructing the
whole inference mechanism. This is not admitted in expectation inference: all
statements are allowed to be revised apart from tautologies (or whatever is a
consequence of the empty set under the underlying  entailment).

In~\cite{MG91} and~\cite{MAK94}, it was observed that preferential entailment
satisfies a weaker form of consistency preservation: there exists a consequence
operation $\vd'$ with $\vd\sbs\vd'\sbs\ev$ such that $\ev$
satisfies Consistency Preservation with respect to $\vd'$. This was proved by
semantical arguments.

In the following theorem, we  make  this property more precise by
expressing it in syntactic terms. We  show that the required underlying
consequence operation retains the properties of the initial one, as it only
differs on the set of assumptions. Therefore, the relation between an expectation  and
a rational inference relation is that of a logic with its theory.

For the proof of Theorem~\ref{thm:consistency-preservation}, we shall make use of the
following  rules (derived in any preferential inference relation).

\begin{lemma}\label{lem:addrules}
In any preferential inference relation, the following rules hold
\begin{enumerate}
\item
$\displaystyle \frac{\ga\ev\gb\quad\gb\ev\ga\quad\ga\ev\gc}{\gb\ev\gc}$
{\rm\small(Reciprocity)}\label{rule:reciprocity}
\vspace{.15in}

\item
$\displaystyle \frac{\ga\land\gb\ev\gc}{\ga\ev\gb\to\gc}$ {\rm\small(S)}
\label{rule:S}
\vspace{.15in}

\item
$\displaystyle \frac{\ga\ev\bot}{\ga\land\gb\ev\bot}$\label{rule:botand}
\vspace{.15in}

\item $\displaystyle \frac{\ga\land\gb\ev\bot}{\ga\ev\neg\gb}$
\label{rule:andbot}
\vspace{.15in}

\item $\displaystyle \frac{\ga\lor\gb\ev\bot}{\ga\ev\bot}$
\label{rule:orbot}
\end{enumerate}
\end{lemma}

\begin{proof}
Rules~\ref{rule:reciprocity},~\ref{rule:S} and~\ref{rule:botand} were introduced and
shown to be derived in a preferential relation in~\cite{KLM90}.
For~\ref{rule:andbot}, suppose
$\ga\land\gb\ev\bot$. Applying  S, we get $\ga\ev\gb\to\bot$ and, by Right weakening,
we conclude~\ref{rule:andbot}. For~\ref{rule:orbot}, suppose
$\ga\lor\gb\ev\bot$. Then, by~\ref{rule:botand}, we get $(\ga\lor\gb)\land\ga\ev\bot$
and, by Left Logical Equivalence, we conclude~\ref{rule:orbot}.\qed
\end{proof}

\begin{theorem}\label{thm:consistency-preservation}
Let $\ev$ be a preferential inference relation based on $\vd$. Then $\ev$ is  a
preferential inference relation based on $\vd'$ that satisfies the Consistency
Preservation rule, where
$$\ga\vd'\gb\qquad \hbox{iff}\qquad \Gamma,\ga\vd\gb,$$
and
$$\Gamma=\{\neg\gc : \gc\ev\bot\}.$$
\end{theorem}

\begin{proof}
We must prove that $\ev$ satisfies Supraclassicality, Left Logical
Equivalence, Right Weakening and Consistency Preservation with respect to
$\vd'$. The rest of the rules are already satisfied since they do not involve an
underlying consequence relation.

First notice that Consistency Preservation is immediate by definition of $\vd'$.

For Supraclassicality, suppose $\ga\vd'\gc$ then $\GC,\ga\vd\gc$. By compactness
of $\vd$, there exist $\gb_1,\ldots,\gb_n \in \cl$ such that
$\gb_1\ev\bot,\ldots,\gb_n\ev\bot$ and $\neg\gb_1,\ldots,\neg\gb_n,\ga\ev\gc$. By
repeated applications of Or, we get $\gb_1\lor\cdots\lor\gb_n\ev\bot$. Let
$\gb=\gb_1\lor\cdots\lor\gb_n$, then
$\gb\ev\bot$ and
$\ga\land\neg\gb\vd\gc$. By Supraclassicality of $\ev$ on $\vd$, we have
$\ga\land\neg\gb\ev\gc$. By Lemma~\ref{lem:addrules}.\ref{rule:botand}, we have
$\ga\land\gb\ev\bot$, so, by Lemma~\ref{lem:addrules}.\ref{rule:andbot}, we have
$\ga\ev\neg\gb$. Using Cut, we get $\ga\ev\gc$, as desired.

For Left Logical Equivalence, suppose $\ga\ev\gc$, $\ga\vd'\gb$, and $\gb\vd'\ga$,
i.e. $\GC,\ga\vd\gb$ and $\GC,\gb\vd\ga$. By compactness, there exist
$\gd_1,\gd_2\in\cl$ such that $\ga\ev\neg\gd_1$, $\ga\ev\neg\gd_2$,
$\ga\land\gd_1\vd\gb$, and $\gb\land\gd_2\vd\ga$. As above, we have $\ga\ev\gb$
and $\gb\ev\ga$. Therefore, by Lemma~\ref{lem:addrules}.\ref{rule:reciprocity},
we get $\gb\ev\gc$, as desired.

Coming to Right Weakening, suppose $\ga\ev\gb$ and $\gb\vd'\gc$, i.e. there
exists $\gd\in\cl$ such that $\ga\ev\neg\gd$ and $\gb\land\neg\gd\ev\gc$. By And,
we have $\ga\ev\gb\land\neg\gd$, so, using Right Weakening of $\ev$ on $\vd$, we
get $\ga\ev\gc$, as desired.
\qed
\end{proof}

Notice that the result applies to rational inference relations, as well, since the
latter are preferential, by definition.
We interpret the above result as follows. Once we
strengthen the underlying entailment, rational inference will become an
expectation inference and, therefore, can be treated as such. It also
implies that the logic of hard and soft constraints is basically the same, their only
difference being what we consider a  consequence of the underlying propositional
entailment. Hard constraints are just taking a place in our belief set as ``guarded''
as that of, say, tautologies. Whatever remains is subject to revision, and hence a
soft constraint.

\section{Rational inference and orderings}
\label{sec:orderings}

Now that we established the correspondence between rational and expectation
inference relations, we shall extend it to a particularly attractive
characterization of the latter with orderings of formulas. We shall first review
G\"ardenfors-Makinson's results and then present our own.

The intuition behind ordering-based formalisms is common in works  on belief
revision, possibilistic logic, and decision theory. We order sentences according to
our expectations. A relation ``$\ga<\gb$'' is interpreted as ``$\gb$ is expected
more than $\ga$'', or ``$\ga$ is more surprising than $\gb$'', or ``$\gb$ is
more possible than $\ga$''.  One can treat such an ordering as a primary
notion; this is the approach of this paper. However, in case of rational
orderings, one can show that such an ordering induces a function from the
extensions of formulas to the unit interval. This function  induces a
{\em possibility } measure on the extensions of formulas (see \cite{DUB86}). A
possibility measure is a ``weak'' probability measure on these extensions.
Roughly, it replaces addition with maximisation. Although the connections with
probability are not clear yet (see
\cite{DUP88}, \cite{SPO87}), probability measures seem especially suited for modeling
cases under uncertainty. Further, a possibility measure arises naturally out of a
database. Zadeh's theory for approximate reasoning (\cite{ZAD78b}) provides a method
for turning available information of a certain form (``fuzzy'' database) into a
possibility measure and, therefore, gives rise to a rational ordering of sentences.

We find that, by a logical point of view, orderings correspond to  prioritization.
We prefer a proof-theoretic  reading, made more explicit in Section~\ref{sec:ranked},
``$\ga$ is more defeasible than $\gb$'' or ``$\ga$ has lower priority  than $\gb$''. A
notion of proof is developed in Section~\ref{sec:ranked} based on this prioritization
and justifies the use of rational orderings without appealing to some probabilistic
intuition.

\begin{definition}{\cite{GM94}}
A {\em rational ordering\/}  is a relation $\leq$ on $\cl$ which
satisfies the following properties:

\begin{tabular}{ll}
1. If $\ga\leq\gb$ and $\gb\leq\gc$, then $\ga\leq\gc$ & (Transitivity), \\
2. If $\ga\vd\gb$, then $\ga\leq\gb$ & (Dominance), \\
3. $\ga\leq\ga\land\gb$ or $\gb\leq\ga\land\gb$ & (Conjunctiveness).
\end{tabular}
\end{definition}

The original name of these orderings was expectation
orderings. However, we shall see that this name is not justifiable, since expectation
inference relations correspond to a smaller class of orderings (see
Definition~\ref{def:expectation-orderings}).

One can easily derive from the above
properties that a rational ordering satisfies
\begin{enumerate}
\item {\em connectivity\/}, i.e.
$\ga\leq\gb$ or $\gb\leq\ga$, and
\item either $\ga\leq\gb$, for all $\gb\in\cl$,
or $\neg\ga\leq\gb$, for all $\gb\in\cl$.
\end{enumerate}

We should mention that the above properties of rational relations are not
new. It is not easy to assign credits, but they have appeared in works in
belief revision (\cite{GM94}), possibilistic logic (\cite{DUP88}),
\cite{DUP91}), fuzzy logic (\cite{ZAD78}), theory of evidence
(\cite{SHAF76}), and economics (\cite{SHAC61}) (see \cite{GM94} for a
historical reference).

G\"ardenfors and Makinson define the following maps between the class of expectation
inference relations and rational orderings.

\begin{definition}{\rm \cite{GM94}}
Given a rational ordering $\leq$ and an expectation inference relation $\ev$, then
define a consequence relation $\ev'$ and an ordering $\leq'$ as follows\\
\begin{tabular}{lrcl}
($C$) & $\ga\ev'\gc$ & iff & either $\ga\vd\gc$\\
         &            &    & or there is a $\gb\in\cl$ such that
                            $\ga\land\gb\vd\gc$ and $\neg\ga<\gb$.\\
($O$) & $\ga\leq'\gb$ & iff & either $\vd\ga\land\gb$ or
$\neg(\ga\land\gb)\notev\ga$.
\end{tabular}\\
We shall also denote $\ev'$ and $\leq'$ with $C(\leq)$ and $O(\ev)$, respectively.
\end{definition}

Condition ($O$) is critical and due to Rott (\cite{ROT91b}). Now, one can prove the
following.

\begin{theorem}{\rm \cite{GM94}}\label{thm:compexp}
Given a rational ordering $\leq$ and an expectation inference relation $\ev$, then
$C(\leq)$ is an expectation inference relation and $O(\ev)$ is a rational ordering.
Moreover, we have $\ev=C(O(\ev))$.
\end{theorem}

This theorem, although it exhibits the first connection between some class of
nonmonotonic consequence relations and orderings of formulas, has two
disadvantages. First, the way it
achieves consistency preservation is ad hoc. If that was not  the case, then the
condition ($C$) would be inappropriate, since, in the first part, it refers
explicitly to the underlying entailment\footnote{However, the second part should
remain the same since we do not mind having a few more consequences, as long as, the
rules which govern the underlying entailment do not change.}. Second, it fails to show
an isomorphism between the class of expectation inference relations and that
of rational orderings, that is $\leq=O(C(\leq))$.  If the second was not the case, then
the condition ($O$) would use only the expectation inference relation to construct the
ordering. Consider the following example.

\begin{example}
Let $\cd_1=\{\{\top\},\{\ \top,\ga\}\}$ and $\cd_2=\{\{\ga\}\}$.
Now define orderings on
$\cl$ as follows
$$\gb\leq_1\gc\quad\hbox{iff}\quad A\vd\gb\ \hbox{implies}\ A\vd\gc,\ \hbox{for all}\
A\in\cd_1.$$
Similarly, for $\cd_2$ and $\leq_2$. We have that $\leq_1\not=\leq_2$, and by
Proposition~\ref{prop:rank-to-rational-order}, the orderings are rational. However,
they generate the same expectation inference relation, using ($O$).
\end{example}

Drawing from the above intuitions and Theorem~\ref{thm:consistency-preservation}, we
define

\begin{definition}
Given a rational ordering $\leq$ and a rational inference relation $\ev$, then
define a consequence relation $\ev'$ and an ordering $\leq'$ as follows

\begin{tabular}{lrcl}
($\bf C$) & $\ga\ev'\gc$ & iff & either  $\gb\leq\neg\ga$, for all $\gb\in\cl$,\\
         &            &    & or there is a $\gb\in\cl$ such that
                            $\ga\land\gb\vd\gc$ and $\neg\ga<\gb$.\\
($\bf O$) & $\ga\leq'\gb$ & iff & either $\neg(\ga\land\gb)\ev\bot$ or
$\neg(\ga\land\gb)\notev\ga$.
\end{tabular}\\
We shall also denote $\ev'$ and $\leq'$ with ${\bf C}(\leq)$ and ${\bf O}(\ev)$,
respectively.
\end{definition}

For Theorem~\ref{thm:first-representation},
we  need the following lemma.

\begin{lemma}\label{lem:ev-to-order}
Let $\leq$ and $\ev$ be a rational ordering and inference relation, respectively. Then
\begin{enumerate}
\item If $\ga\ev\bot$ then $\gb\leq'\neg\ga$, for all $\gb\in\cl$, where $\leq'={\bf
O}(\ev)$.\label{botev}
\item If $\gb\leq\neg\ga$, for all $\gb\in\cl$, then $\ga\ev'\bot$, where $\ev'={\bf
C}(\leq)$.\label{botord}
\item $\neg\ga < \ga\lto\gc$ iff $\{\gb : \neg\ga<\gb\} \vd \ga\lto\gc$.\label{rest}
\end{enumerate}
\end{lemma}

Now, everything falls into place.

\begin{theorem}\label{thm:first-representation}
Given a rational ordering $\leq$ and a rational inference relation $\ev$, then
${\bf C}(\leq)$ is a rational  inference relation and ${\bf O}(\ev)$ is a rational
ordering. Moreover, we have $\ev={\bf C}({\bf O}(\ev))$ and $\leq={\bf O}({\bf
C}(\leq))$.
\end{theorem}

Now, if rational orderings are in adjunction with rational inference relations, what is
the class of orderings which corresponds to expectation inference relations? For that,
observe that by Lemma~\ref{lem:ev-to-order}, hard constraints are positioned on the
top of rational orderings. So, it is enough to keep exclusively this place for the
consequences of the empty set and add this as a condition to rational orderings.

\begin{definition}
\label{def:expectation-orderings}
An {\em expectation ordering\/}  is a rational ordering  which
satisfies, in addition, the following property:
\begin{quote}
 If $\gb\leq\ga$, for all $\gb\in\cl$, then $\vd\ga$.
\end{quote}
\end{definition}

Now, using the {\em same\/} defining conditions ($\bf C$) and
($\bf O$),  we can state the improved
characterization theorem for expectation inference relations.

\begin{theorem}\label{thm:expectation-orderings-to-inference}
Given an expectation ordering $\leq$ and an expectation inference relation $\ev$, then
${\bf C}(\leq)$ is an expectation inference relation and ${\bf O}(\ev)$ is an
expectation ordering. Moreover, we have $\ev={\bf C}({\bf O}(\ev))$ and $\leq={\bf
O}({\bf C}(\leq))$.
\end{theorem}

Theorems \ref{thm:first-representation} and
\ref{thm:expectation-orderings-to-inference} can now be used for giving new
straightforward proofs for the characterization of rational inference with ranked
preferential models (Theorem~3.12 of
\cite{LM92}) and  expectation inference with  nice preferential models (Theorems
3.8 and 3.9 of \cite{GM94}). These proofs  will
appear elsewhere.

\section{Ranked consequence operators}
\label{sec:ranked}

First, a word about the plan of this section. We  introduce the notion of ranked
consequence operation without referring to an underlying entailment
(Definition~\ref{def:set-ranked}). The reason for such a definition is that we can
motivate ranked consequence operators independently of nonmonotonic reasoning.
Then we define a smaller class based on an underlying entailment
(Definition~\ref{def:rankvd}) and show that this class characterize rational
inference relations. The same constraints we assumed   for a
language
$\cl$ and an entailment $\vd$  in
Section~\ref{sec:preliminaries} continue to hold here.

Think of a reasoner whose beliefs are ordered accordingly to their defeasibility.
Beliefs which are less likely to be defeated come before beliefs which are more
likely to be defeated. For instance ``Birds fly'' will come after ``Penguins
do not fly'' (since the former has more exceptions) and ``Mary is married'' might
come before ``Mary is married with children'' (since the latter is stronger).
There is a natural way to attach a consequence operator to this belief prioritization.

\begin{definition}\label{def:set-ranked}
Let $\langle  I,<\rangle $  be a linear ordering, and
$\{ A_i\}_{i\in I}$ be an upward chain of sets of
formulas  such that $A_i\sbs A_j$ iff $i\leq j$.
Define the following consequence
operators (one for each $i\in I$):
\begin{center}
\begin{tabular}{rcl}
$\ev_i\beta$ &\quad iff & \quad$\gb\in A_i $, \\
$\alpha\ev_i\beta$ &\quad iff &\quad $\neg\alpha\not\in A_i $ and
                           $\alpha\to\beta\in A_i$.
\end{tabular}
\end{center}

Now let

\begin{center}
\begin{tabular}{rcl}

$\alpha\ev\beta$ & \quad iff & \quad either $\alpha\ev_i\beta$, for some $i\in I$,\\
 &  &\quad or $\ev_i\neg\ga$, for all $i\in I$.
\end{tabular}
\end{center}

The consequence operator $\ev$ will be called {\em ranked consequence} operator
(induced by
$\{A_i\}_{i\in I}$).
\end{definition}

First, note that $A_i$'s are {\em not\/} necessarily deductively closed. Second,
notice that,  unless  we add the last part of the definition of $\ev$, we do not
provide  for formulas $\ga$, where $\neg\ga\in A_i$, for all $i\in I$.
In order to have
$\ga\ev\bot $, there must either be an $A_i$ such that $\neg\ga\not\in A_i$ and
$\ga\to\bot\in A_i$, or  $\neg\ga\in A_i$, for all $i\in I$. This means that if our
beliefs can accommodate a context where $\ga$ holds, then we use the part of
 the ordering that remains consistent after adding $\ga$.  Therefore
the $A_i$'s which contain both $\neg\ga$ and
$\ga$ are irrelevant to the consequence operator.

Indices assign grades of relying on
the set of consequences as the next example, formalizing omniscience,
shows.
\bigskip

\begin{example}
 Let $\langle\omega,<\rangle$ be the set
of natural numbers with the usual order. Now let
$\vdash$ be the classical consequence relation and let
$A_1$ be some set of formulas of propositional logic. Let
$$A_2=\{ \phi \mid\ \hbox{is provable in  one step from}\ A_1 \},$$
and, inductively,
$$A_n=\{ \phi \mid\ \hbox{is provable in less than $n-1$ steps from}\  A_1
\}.$$

Notice that if $A_1$ is consistent and $\ev$ is the ranked consequence
operator   defined through
$\{A_i\}_{i\in \omega}$ then
$$\alpha\ev\beta\quad\hbox{iff}\quad A_1,\alpha\vdash\beta,$$
where $\vdash$ is the classical consequence operator of propositional
calculus. Note that if $A_1$ is inconsistent, then $\ga$ entails all formulas
which are provable from $A_1$ with less steps than $\neg\ga$, i.e., before we
realize inconsistency.  Now, if
$A_1$ is the set of all tautologies or, better, an axiomatization of  them then
$\ev$ is exactly the classical consequence operator.
\end{example}

It is clear that the above representation is syntax-based, i.e. depends on the
particular representation of $A_i$'s. The case where the sets of formulas $A_i$ are
closed under consequence is the  one we shall deal with in this paper. Doing that
is like being logically omniscient;  we do not assign any cost to derivations using
$\vd$.

We shall now give a definition of
ranked consequence operator using an underlying entailment.
In case the $A_i$'s are closed under
consequence, it coincides with the original definition  (by replacing the set
belonging relation $\in$ with proposition entailment $\vd$).

\begin{definition}
\label{def:rankvd}
A {\em ranked consequence operator $\ev$ based on $\vd$\/} induced by a chain of
sets
$\{ B_i\}_{i\in I}$ under inclusion is defined as
follows:

We first define a set of consequence operators $\ev_i$ (one for each $i\in I$):

\begin{center}
\begin{tabular}{rcl}
$\ga\ev_i\gb$ &\quad iff & \quad $B_i\not\vd\neg\ga$ and $B_i,\ga\vd\gb$.
 \end{tabular}
\end{center}
Note that we denote  $\top\ev_i\gb$ with $\ev_i\gb$. We can now let
\begin{center}
\begin{tabular}{rcl}
$\ga\ev\gb$ &\quad iff & \quad either $\ga\ev_i\gb$, for some $i\in I$,\\
            &       & \quad or $\ev_i\neg\ga$, for all $i\in I$.
\end{tabular}

\end{center}
We shall use $\langle \{ B_i \}_{i\in I},\vd\rangle$ to denote this operator.
\end{definition}

Notice that we can have both $\not\vd\ga$ and  $\ev_i\neg\ga$, for
all $i$. This translates to the fact that $\ga$ can be true in some possible world
but it is unthinkable for us to include it in our beliefs.  The above mechanism treats
such a case as an instance of a {\em hard\/} constraint: such an $\ga$ implies
falsehood.

Again notice that unless $B_i\vd \neg\ga$, for all $i \in I$, we cannot derive
falsity from $\ga$. The reason is that, in those cases, we are able to form a
context based on $\ga$ (a chain of sets of formulas which prove $\ga$)  which
is consistent. Again, the inconsistent $B_i$'s   are irrelevant to the
consequence operator. The following proposition allows us to assume that the ordering
is complete, that is it has all meets and joins, and amounts essentially to Lewis'
assumption or smoothness property of preferential models.

\begin{proposition}
\label{prop:complete}
A  ranked consequence operator $\ev$ based on $\vd$
is induced by a chain of sets of formulas if and only if it is induced by
the closure of this chain under arbitrary unions and intersections.
\end{proposition}

This result has the following significance: it allows an assignment of a rank to
an assertion of  the form $\ga\ev\gb$. Suppose that $\ga\ev\gb$ holds. If
$\ev_i\neg\ga$ for all $i\in I$ does not hold, then the set
$I_{\ga\ev\gb}=\{i:\notev_i\neg\ga\hbox{ and }\ga\ev_i\gb\}$ is not empty. Moreover,
it is  connected. Now, it is easy to see that,  in the completion of the chain, this
set has a least element (because it is closed under intersection) and  a greatest
element (because it is closed under unions). Let $i_1$ and $i_2$ be the
indices of the least and greatest elements, respectively. The {\em rank\/} of the
assertion $\ga\ev\gb$ is $i_1$ and its {\em range \/} $[i_1,i_2]$. In case
$\ev_i\neg\ga$, for all $i\in I$, then set the rank of $\ga\ev\gb$ to
$0$ and its range to $[0,l]$, where $0$  and $l$ are the indices of least and the
greatest element of the linear order, respectively. In case of an assertion
$\ev\gb$,  observe that its range is of the form $[i,l]$, where $l$ is the index of
the greatest element of the linear ordering.

Finally, notice that a ranked consequence operator is not necessarily monotonic.

\begin{example}
Let $B_1=\{\ga\}$ and $B_2=\{\ga,\neg\gb\}$. We have $\ga\ev\neg\gb$ because
$\neg\ga\not \in Cn(B_2)$ and $\ga\to\neg\gb\in Cn(B_2)$. But we also have that
$\ga\land\gb\notev\neg\gb$ because $\neg(\ga\land\gb)\not\in Cn(B_1)$.
\end{example}

Now, it is interesting to ask what kind of properties a ranked consequence operator
satisfies. It turns out that each ranked consequence operator gives rise to a
rational inference relation. Although one can show it directly, we  define the
rational orderings induced by such operators.

\begin{definition}\label{def:ranked-consequence-operator}
Given a ranked consequence operator, let
$$\ga\leq\gb\quad \hbox{iff} \quad B_i\vd \ga\ \hbox{implies}\  B_i\vd \gb,\ \hbox{for all
$i\in I$}.$$
and call $\leq$ the {\em ordering induced by the ranked consequence operator $\ev$\/}.
\end{definition}

We, now, have the following

\begin{proposition}\label{prop:rank-to-rational-order}
An ordering induced by a ranked consequence operator is rational. Moreover, ${\bf
C}(\leq)=\ev$.
\end{proposition}

We have  immediately the following.

\begin{corollary}\label{corol:rank-to-rational}
A ranked consequence operator is a rational inference relation.
\end{corollary}


The other direction of the above theorem holds, too. We should only show, given a
rational ordering, how to generate a total order of sets of formulas.
To this end, we shall define
a chain of sets $\{A_i\}_{i\in I}$ which generates a ranked consequence
operator $\ev$ equal to ${\bf C}(\leq)$. Let $\sim$ be the equivalence relation
induced by $\leq$ (a rational ordering is  a preorder). The
equivalence classes will be denoted by $\hat{\ga}$ (where $\ga\in\hat{\ga}$). It
is also clear that the set of equivalence classes is linearly ordered.
Now, for each $\ga\in\cl$,  let
$$A_{\hat{\ga}}=\{\gb : \ga\leq\gb\}.$$
Note here that, by Dominance, the sets $A_{\hat{\ga}}$ are closed under
consequence. Moreover, we have $A_{\hat{\ga}}\sbs A_{\hat{\gb}}$ iff $\gb\leq\ga$.
Now, generate a ranked consequence operator
$\ev$ as in Definition~\ref{def:ranked-consequence-operator}.
This ranked consequence operator turns out to
be equal to the one generated by the rational order.
So, we have the following.

\begin{theorem}\label{thm:rational-order-to-rank}
A  rational inference relation is a ranked consequence operator.
\end{theorem}

The proof of the above theorem shows that a rational ordering can be defined
by  a chain of sets which induces a ranked consequence operator and conversely.
However,  the same  rational ordering can be induced by two
different ranked consequence operators.  This should hardly be surprising, as ranked
consequence operators play the role of axiomatizing a nonmonotonic ``theory'', that
is, a rational inference. Moreover,
\begin{itemize}
\item ranked consequence operators are proof-theoretic in their motivation, and
therefore closer to what we want to describe by a rational inference relation, and
\item a ranked consequence operator assigns ranks to assertions as well as to
formulas therefore grading the whole process of inference.
\end{itemize}

We showed that  rational and expectation inference
relations are exactly the same class of consequence relations if we allow the
underlying propositional entailment to ``vary''. However fixing
$\vd$, is it possible to tell if a ranked consequence operator satisfies
Consistency Preservation? The answer is affirmative, for a formula $\ga$ infers
inconsistency ($\ga\ev\bot$) if and only if its negation is a consequence of the
first element of the chain which induces the ranked consequence operator (as a
corollary of Proposition~\ref{prop:complete}, a first element always exists). To see
that, suppose
$\ga\ev\bot$, then, by definition, we must have
$\ev_i\neg\ga$, for all
$i \in I$, and for that it is enough that the first element of the chain implies
$\neg\ga$.

\section{Rational default systems}
\label{sec:rational-default}

In this section, we shall see how one can design a ranked consequence operator.
Suppose we are given a  number of sets of (normal, without prerequisites)
defaults in a linear {\em well-founded\/} prioritization. Moreover, and this is an
important  assumption for rational inference, we are asked to, either apply the whole
set, or not apply it at all. Let the set of sets of defaults be
$\cd=\{A_i\}_{i\in I}$, where $\langle I,<\rangle$ is a well-founded linear strict
order, and $A_i$ is preferred from $A_j$ whenever $i<j$. The are two ways to read this
preference.

\begin{itemize}
\item The first way is a {\em strict\/} one: if you cannot add $A_i$ to
your set of theorems (that is, you derive inconsistency by adding $A_i$) then
you cannot add $A_j$, for all $A_j$ less preferred from $A_i$.
\item the other is {\em liberal\/}: if you cannot add $A_i$ to your set of
theorems, then you can add $A_j$, where $A_j$ is less preferred from $A_i$, {\em
provided\/} you cannot add $A_k$, where $A_k$ is more preferred than $A_j$.
\end{itemize}

To illustrate this, consider the following example.

\begin{example}
Let $\cd=\{A_1,A_2,A_3\}$, where $A_1=\{\ga\lto\gb\}$, $A_2=\{\neg\gb\}$, and
$A_3=\{\gb\lto\gc\}$. Assume $\ga$. Following the strict interpretation,
we can only infer
$\gb$, from $A_1$. With the liberal interpretation, we can also infer $\gc$,
since we are allowed to add $A_3$, and cannot add $A_2$ that leads to a
contradiction.
\end{example}

It turns out that those readings are equivalent. Not in the sense that the {\em
same\/} set of sets of defaults generate the same consequences, but that a
strict extension of a family of sets of defaults can be reduced to a liberal
extension of another family of sets of defaults, and conversely. It can be easily
shown that strict and liberal extensions of families of sets of defaults are
instances of rational consequence operators and, therefore, rational. In particular,
 Proposition~\ref{prop:rank-to-rational-order} gives us a way to
construct the rational orderings of such default systems.

Given $\cd=\{A_i\}_{i\in I}$, where $\langle I, < \rangle$ is well-founded, define the
following strict  ordering between non-empty subsets of $I$:

\begin{tabular}{rcll}
$K<L$ & iff & there exists $i\in I$ such that & a. $i\in K$ but $i\not\in L$,
                                                      and\\
       &   &              & b. for all $j<i$, $j\in K$ iff $j\in L$.
\end{tabular}

It can be shown that $<$ is linear. Now, let
$A_K=\bigcap_{K\leq L} Cn(\bigcup_{i\in L} A_i)$.

\begin{definition}
Let $\ga\in\cl$ and $\cd=\{A_i\}_{i\in I}$, where  $\langle I,<\rangle$ is a
total strict order, and $A_i\sbs\cl$, for all $i\in I$. The {\em strict extension
$E^s_\cd(\ga)$ of $\ga$ with respect to $\cd$\/} is defined as follows
$$E^s_\cd(\ga)=\bigcup\{E^s_i : \hbox{$E^s_i$ is consistent}\},$$
where $E^s_i=Cn(\{\ga\}\cup\bigcup_{j\leq i}A_j)$.
The {\em liberal extension
$E^l_{\cd}(\ga)$ of $\ga$ with respect to $\cd$\/} is defined as follows
$$E^l_{\cd}(\ga)=\bigcup\{E^l_L : \hbox{$E^l_L$ is consistent}\},$$
where $E^l_L=Cn(\{\ga\}\cup\bigcup_{K\leq L}A_K)$.
\end{definition}

Thus the liberal extension of $\cd=\{A_i\}_{i\in I}$ is the strict extension of
$\cd'=\{A_K\}_{K\in \cp(I)^*}$. For the other direction, the strict extension of
$\cd=\{A_i\}_{i\in I}$ coincides with the liberal extension of
$\cd''=\{C_i\}_{i\in I}$, where $C_i=\bigcup_{j\leq i}A_i$.

So, it is enough to construct the ranked consequence operator for the strict
extension of $\cd=\{A_i\}_{i\in I}$. But this is easily achieved. Consider
$\langle \{B_i\}_{i\in I},\ev\rangle$, where $B_i=E^s_i$.

Thus, strict and liberal extensions of prioritized sets of set of formulas are
rational. The above definition,
together with Proposition~\ref{prop:rank-to-rational-order}, gives us a way to
construct the rational orderings of such default systems. Given a prioritized set
$\cd=\{A_i\}_{i\in I}$ then the rational ordering of its strict extension is
$$\ga\leq^s_{\cd}\gb\quad \hbox{iff}\quad \bigcup_{j\leq i}A_j\vd\ga\
\hbox{implies}\ \bigcup_{j\leq i}A_j\vd\gb,\ \hbox{for all $i\in I$}.$$
The
rational ordering of its liberal extension is
$$\ga\leq^l_{\cd}\gb\quad \hbox{iff}\quad \bigcup_{K\leq L}A_K\vd\ga\
\hbox{implies}\ \bigcup_{K\leq L}A_K\vd\gb,\ \hbox{for all}\ L\in \cp(I)^*.$$

Assuming a finite language, sets of formulas, intersections, and
unions of them correspond to conjunctions, disjunctions, and conjunctions,
respectively.


 A study of the above
default systems under the assumption of finite language, has been carried already in
the context of belief revision (therefore, assuming consistency preservation, in
addition to finite language), by Nebel (\cite{NEB91},
\cite{NEB94}). Our strict and liberal extensions are called {\em prioritized\/}
and {\em linear base revision\/}, respectively.  Also,  Nebel showed
in~\cite{NEB94} that  deciding if a certain formula is contained in the strict
or in a liberal extension (that is, deciding
$\ga\ev\gb$) is
$\mathrm{P}^{\mathrm{NP} [O(\log^c n)]}$ and $\mathrm{P}^{\mathrm{NP}
[O(n)]}$, respectively. We expect that these results carry over to our framework.

\section{Conclusion}
\label{sec:conclusion}

We summarize  our results  in the following
\begin{theorem}
Let $\ev$ be a binary relation on $\cl$. Then the following are equivalent:

\begin{enumerate}
\item $\ev$ is a rational inference relation, i.e. it satisfies Supraclassicality,
Left Logical Equivalence, Right Weakening, And, Cut, Cautious Monotonicity, Or, and
Rational Monotonicity.
\item $\ev$ is characterized by some rational relation $\leq$ on $\cl$ using
condition ($\bf C$).
\item $\ev$ is defined by a ranked consequence operator.
\end{enumerate}
\end{theorem}

Since rational orderings are in one-to-one correspondence with rational
inference, our first representation result has many ramifications. Results
in belief revision can be translated in a nonmonotonic framework and vice versa.
For instance, selection functions and preferential models can be used for the
modeling of both. Proofs of this results are straightforward through our defining
conditions for rational and expectation orderings. Work that has already been
done on expectation inference relations (e.g. the study of generating expectation
inference relations through incomplete rational orderings---see \cite{FHL94})
can be lifted smoothly to rational inference.

Our second representation result reveals the working mechanism of rational
inference. It shows that, in order to attain rational inference, we must
prioritize defaults in a particular way. We showed how default logic formalisms
can fit this pattern. It  enables us to assign grades to all components of the
reasoning system (formulas and rules). Therefore, it is a particular attractive
way to use it as an inference mechanism for nonmonotonic reasoning.

The above  characterization results reveal
another notion of consequence paradigm hidden behind nonmonotonicity. However, apart
from Dubois and Prade's work on possibility logic, this paradigm has been passed
largely unrecognized by logicians as an appropriate  method for a treatment of
vagueness and uncertainty. Yet, this paradigm arose independently from various studies
on different fields and appeared  before nonmonotonic logic. In addition, it is
applicable. Now, an important question arises: how far this paradigm extends. In other
words, is it possible to reduce a nonmonotonic consequence relation to some relation
expressing prioritization? The answer is positive and uniform. The
important case of preferential consequence relations is treated separately in
\cite{KG95e}, while the general case (which includes cumulative consequence relations)
appears in \cite{KG96f}.
\medskip

\noindent {\bf Acknowledgements:} I would like to thank Gianni Amati and R.
Ramanujam for their helpful comments on a preliminary version of this paper.

\newpage

\appendix
\section{Proofs}

For the proof of Theorem~\ref{thm:consistency-preservation}, we shall make use of the
following  rules (derived in any preferential inference relation).

\begin{alemma}{\bf \ref{lem:addrules}}\quad
In any preferential inference relation, the following rules hold
\begin{enumerate}
\item
$\displaystyle \frac{\ga\ev\gb\quad\gb\ev\ga\quad\ga\ev\gc}{\gb\ev\gc}$
{\small(Reciprocity)}

\item
$\displaystyle \frac{\ga\land\gb\ev\gc}{\ga\ev\gb\to\gc}$ {\small(S)}

\item
$\displaystyle \frac{\ga\ev\bot}{\ga\land\gb\ev\bot}$

\item $\displaystyle \frac{\ga\land\gb\ev\bot}{\ga\ev\neg\gb}$

\item $\displaystyle \frac{\ga\lor\gb\ev\bot}{\ga\ev\bot}$

\end{enumerate}
\end{alemma}

\begin{proof}
Rules~\ref{rule:reciprocity},~\ref{rule:S} and~\ref{rule:botand} were introduced and
showed to be derived in a preferential relation in~\cite{KLM90}.
For~\ref{rule:andbot}, suppose
$\ga\land\gb\ev\bot$. Applying  S, we get $\ga\ev\gb\to\bot$, and, by Right weakening,
we conclude~\ref{rule:andbot}. For~\ref{rule:orbot}, suppose
$\ga\lor\gb\ev\bot$. Then, by~\ref{rule:botand}, we get $(\ga\lor\gb)\land\ga\ev\bot$,
and, by Left Logical Equivalence, we conclude~\ref{rule:orbot}.
\end{proof}

\begin{atheorem}{\bf \ref{thm:consistency-preservation}}\quad
Let $\ev$ be a preferential inference relation based on $\vd$. Then $\ev$ is a
preferential inference relation based on $\vd'$ that satisfies the Consistency
Preservation rule, where
$$\ga\vd'\gb\qquad \hbox{iff}\qquad \Gamma,\ga\vd\gb,$$
and
$$\Gamma=\{\neg\gc : \gc\ev\bot\}.$$
\end{atheorem}

\begin{proof}
We must prove that $\ev$ satisfies Supraclassicality, Left Logical
Equivalence, Right Weakening and Consistency Preservation with respect to
$\vd'$. The rest of the properties are already satisfied since $\ev$ is a
rational inference relation.

First notice that Consistency Preservation is immediate by definition of $\vd'$.

For Supraclassicality, suppose $\ga\vd'\gc$ then $\GC,\ga\vd\gc$. By compactness
of $\vd$, there exists $\gb_1,\ldots,\gb_n \in \cl$ such that
$\gb_1\ev\bot,\ldots,\gb_n\ev\bot$ and $\neg\gb_1,\ldots,\neg\gb_n,\ga\ev\gc$. By
repeated applications of Or, we get $\gb_1\lor\cdots\lor\gb_n\ev\bot$. Let
$\gb=\gb_1\lor\cdots\lor\gb_n$, then
$\gb\ev\bot$ and
$\ga\land\neg\gb\vd\gc$. By Supraclassicality of $\ev$ on $\vd$ we have
$\ga\land\neg\gb\ev\gc$. By Lemma~\ref{lem:addrules}.\ref{rule:botand}, we have
$\ga\land\gb\ev\bot$, so, by Lemma~\ref{lem:addrules}.\ref{rule:andbot}, we have
$\ga\ev\neg\gb$. Using Cut, we get $\ga\ev\gc$.

For Left Logical Equivalence, suppose $\ga\ev\gc$, $\ga\vd'\gb$ and $\gb\vd'\ga$,
i.e. $\GC,\ga\vd\gb$ and $\GC,\gb\vd\ga$. By compactness there exist
$\gd_1,\gd_2\in\cl$ such that $\ga\ev\neg\gd_1$, $\ga\ev\neg\gd_2$,
$\ga\land\gd_1\vd\gb$ and $\gb\land\gd_2\vd\ga$. As above we have $\ga\ev\gb$
and $\gb\ev\ga$. Therefore by Lemma~\ref{lem:addrules}.\ref{rule:reciprocity}
we get $\gb\ev\gc$.

Coming to Right Weakening, suppose $\ga\ev\gb$ and $\gb\vd'\gc$, i.e. there
exists $\gd\in\cl$ such that $\ga\ev\neg\gd$ and $\gb\land\neg\gd\ev\gc$. By And
we have $\ga\ev\gb\land\neg\gd$, so using Right Weakening of $\ev$ on $\vd$ we
get $\ga\ev\gc$.

\end{proof}

For Theorem~\ref{thm:first-representation},
we shall need the following lemma.

\begin{alemma}
{\bf \ref{lem:ev-to-order}}
\quad
Let $\leq$ and $\ev$ be a rational ordering and inference relation, respectively.
Then
\begin{enumerate}
\item If $\ga\ev\bot$ then $\gb\leq'\neg\ga$, for all $\gb\in\cl$,
where $\leq'={\bf O}(\ev)$.
\item If $\gb\leq\neg\ga$, for all $\gb\in\cl$, then $\ga\ev'\bot$, where $\ev'={\bf
C}(\leq)$.
\item $\neg\ga < \ga\lto\gc$ iff $\{\gb : \neg\ga<\gb\} \vd \ga\lto\gc$.
\item If $\ga\ev'\bot$ then $\gb\leq\neg\ga$, for all $\gb\in\cl$,
where $\ev'={\bf C}(\leq)$.\label{ordtbot}
\item If $\gb\leq'\neg\ga$, for all $\gb\in\cl$, then $\ga\ev\bot$,
where $\leq'={\bf O}(\ev)$.\label{evtbot}
\end{enumerate}
\end{alemma}

\begin{proof}
Part~\ref{botord} is immediate from defining condition ($\bf C$).
For Part~\ref{botev}, suppose that $\neg(\neg\ga\land\gb)\ev\gb$. We must show
$\neg(\neg\ga\land\gb)\ev\bot$. Since $\ga\ev\bot$, we have $\ga\ev\neg\gb$, by
Right Weakening. Applying Or, we get $\ga\lor\neg\gb\ev\neg\gb$. By hypothesis and
And, we get $\neg(\neg\ga\land\gb)\ev\bot$.

The left to right direction of  Part~\ref{rest} is straightforward.
For the right to left direction, suppose $\{\gb : \neg\ga<\gb\} \vd \ga\lto\gc$.
Then, by Compactness, there exist
$\gb_1,\ldots,\gb_n$ such that $\neg\ga < \gb_i$, for all
$i\in\{1,\ldots,n\}$, and $\gb_1\land\cdots\land\gb_n\vd\ga\lto\gc$. By
Conjunctiveness, we have $\neg\ga <\gb_1\land\cdots\land\gb_n$. So, by Dominance, we
have $\gb_1\land\cdots\land\gb_n\leq \ga\lto \gc $. Hence, by Transitivity,
$\neg\ga < \ga\lto \gc $, as desired.

For  Part~\ref{ordtbot}, we have, by definition of $\ev'$ that $\gb\leq\neg\ga$, for
all
$\ga\in\cl$ or there is $\gb\in\cl$ such that $\ga\land\gb\vd\bot$ and $\neg\ga <
\gb$. If the former holds then the result follows immediately. If the latter holds
then $\gb\vd\neg\ga$ that contradicts $\neg\ga < \gb$, by Dominance.
\marginpar{Do Part~\ref{evtbot}}
\end{proof}

\begin{lemma}
Let $\leq$ and $\ev$ be a rational ordering and inference relation, respectively.
Then
\begin{enumerate}
\item $\neg\ga\lor\neg\gb\ev\neg\ga$ implies $\ga\leq'\gb$, where $\leq'={\bf
O}(\ev)$. \label{neg-ev-to-leq}
\item $\ga\ev'\gb$ implies $\neg\ga\lor\neg\gb\leq\neg\ga$,
where $\ev'={\bf C}(\leq)$.\label{neg-leq-to-ev}
\item If $\leq$ satisfies Bounded Disjunction then $\neg\ga\lor\neg\gb\leq\neg\ga$
implies $\ga\ev'\gb$, where $\ev'={\bf C}(\leq)$.\label{neg-ev-to-leq-iso}
\end{enumerate}
\end{lemma}

\begin{proof}
For Part~\ref{neg-ev-to-leq}, if $\neg\ga\lor\neg\gb\notev\ga$ we get immediately
$\ga\leq'\gb$, by Definition ({\bf O}). If not, that is $\neg\ga\lor\neg\gb\ev\ga$,
then, by And and Right Monotonicity, we have $\neg\ga\lor\neg\gb\ev\bot$. Again, by
Definition ({\bf O}), we have $\ga\leq'\gb$.

For Part~\ref{neg-leq-to-ev}, assume $\ga\ev'\gb$. If $\ga\vd\gb$.
Then we have $\neg\gb\vd\neg\ga$, so $\neg\ga\lor\neg\gb\vd\neg\ga$, and hence, by
Dominance,  $\neg\ga\lor\neg\gb\ev\neg\ga$, as desired. If not then there must be
$\gc\in\cl$ such that $\gc\vd\ga\lto\gb$ and $neg\ga < \gc$. Therefore $\neg\ga <
\neg\ga\lor\neg\gb$. Now, suppose
$\neg\ga\lor\neg\gb\not\leq\neg\ga$ towards a contradiction. By
Connectivity, we have
$\neg\ga < \neg\ga\lor\neg\gb$ and so, by Conjunctiveness, $\neg\ga <
(\ga\lto\gb)\lor(\ga\lto\neg\gb)$. Hence $\neg\ga < \neg\ga$, a contradiction to
Reflexivity.

For Part~\ref{neg-ev-to-leq-iso}, if $\neg\ga < \ga\lto\gb$ then we immediately have
$\ga\ev'\gb$, by Definition ({\bf C}). If not, that is $\ga\lto\gb\leq\neg\ga$, then
applying Bounded Disjunction, we have $\ga\lor\gb\lor\neg\gb\leq\neg\ga$. The latter
implies $\top\leq\neg\ga$, so, by Dominance $\gc\leq\neg\ga$, for all $\gc\in\cl$.
Hence $\ga\ev'\gb$, by Definition ({\bf C}).
\end{proof}

\begin{atheorem}{\bf \ref{thm:first-representation}}\quad
Given a rational ordering $\leq$ and a rational inference relation $\ev$, then
${\bf C}(\leq)$ is a rational inference relation and ${\bf O}(\ev)$ is a rational
ordering. Moreover, we have $\ev={\bf C}({\bf O}(\ev))$ and $\leq={\bf O}({\bf
C}(\leq))$.
\end{atheorem}

\begin{proof}
We shall try not to overlap with the proof of G\"ardenfors and Makinson proof
of Theorem~\ref{thm:compexp} (see proof of Theorem~3.3 in
\cite{GM94}). Therefore we do not cover the case where    the second
half of condition (R$\ev$) applies. The list of rules we verify is
Supraclassicality, Left Logical Equivalence, And, Cut, Cautious Monotony, Or and
Rational Monotony. Right Weakening follows from the above list.

We shall first show that ${\bf C}(\leq)$ is a
rational inference relation.

For Supraclassicality, suppose that $\ga\vd\gc$ but  not
$\gb\leq\neg\ga$ for all $\gb\in\cl$. So there exists $\gb\in\cl$ such that
$\neg\ga<\gb$. But then
$\ga\land\gb\vd\gc$ and therefore $\ga\ev\gc$.

For Left Logical Equivalence, suppose that $\ga\ev\gc$, $\vd\ga\lto\gb$ and
$\gd\leq\neg\ga$ for all $\gd\in\cl$. Since $\gb\vd\ga$ we have
$\neg\ga\vd\neg\gb$. By Dominance we get
$\neg\ga\leq\neg\gb$ and by Transitivity $\gd\leq\neg\gb$ for all $\gd\in\cl$.
Therefore
$\gb\ev\gc$.

For And, suppose that $\ga\ev\gb$ and $\ga\ev\gc$. In case $\gd\leq\neg\ga$ for
all $\gd\in\cl$ we have immediately $\ga\ev\gb\land\gc$.

Turning to Or, suppose that $\ga\ev\gc$ and $\gb\ev\gc$.  If $\gd\leq\neg\ga$ for
all $\gd\in\cl$ and $\gd\leq\neg\gb$ for
all $\gd\in\cl$, then by Conjunctiveness we have either
$\neg\ga\leq\neg\ga\land\neg\gb$ or $\neg\gb\leq\neg\ga\land\neg\gb$. In
either case $\gd\leq\neg\ga\land\neg\gb$ for all $\gd\in\cl$ by Transitivity.
Therefore $\neg(\ga\lor\gb)\ev\gc$. In the mixed case, say $\gd\leq\neg\ga$ for
all $\gd\in\cl$ and there exists $\gd_0\in\cl$ such that $\gb\land\gd_0\vd\gc$
and $\neg\gb<\gd_0$, we have $(\ga\lor\gb) \land (\neg\ga\land\gd_0) \vd\gc$.
Now suppose that $\neg\ga\land\gd_0 < \gd_0$. By Conjunctiveness we must have
$\gd_0\leq\neg\ga\leq\neg\ga\land\gd_0$, a contradiction. Thus
$\neg(\ga\lor\gb)\leq\gb < \gd_0 \leq \neg\ga\land\gd_0$. Therefore
$\ga\lor\gb\ev\gc$.

For Cut, suppose that $\ga\ev\gb$ and $\ga\land\gb\ev\gc$. If $\gd\leq\neg\ga$ for
all $\gd\in\cl$ then, by definition, $\ga\ev\gc$. If not, there exists
$\gd_0\in\cl$ such that $\ga\land\gd_0\vd\gb$ and $\neg\ga < \gd_0$.  Now
suppose that  $\gd\leq\neg(\ga\land\gc)$ for all $\gd\in\cl$. Observe that
$\ga\land [ ( \neg\ga\lor\neg\gb) \land \gd_0 ] \vd \gc$. We moreover have that
$ \neg\ga < \gd_0 \leq (\neg\ga\lor\neg\gb) \land\gd_0 $. Therefore $\ga\ev\gc$.

For Rational Monotonicity, suppose that $\ga\ev\gc$ and $\ga\notev\neg\gb$.
If $\gd\leq\neg\ga$ for all $\gd\in \cl$, then we get a contradiction because
$\ga\ev\neg\gb$.

For Cautious Monotony, suppose that $\ga\ev\gb$ and $\ga\ev\gc$.
Observe that in case $\ga\notev\neg\gb$ then the result follows by an
application of Rational Monotony. If not, i.e. $\ga\ev\neg\gb$, then by applying
And we have  $\ga\ev\bot$.  If $\gd\leq\neg\ga$ for all $\gd$, then, since
$\neg\ga\vd\neg\ga\lor\neg\gb$, we have
$\gd\leq\neg\ga\leq\neg\ga\lor\neg\gb$ but
$\vd\neg\ga\lor\neg\gb\lto\neg(\ga\land\gb)$ therefore
$\ga\land\gb\ev\gc$. Otherwise there exists $\gd$ such that $\ga\land\gd\vd\bot$
and $\neg\ga<\gd$. But then we have that $\gd\vd\neg\ga$ and therefore
$\gd\leq\neg\ga$ which is a contradiction to our hypothesis.

Definition ($\bf O$) is identical to G\"{a}rdenfors and Makinson's one in the
second disjunct. Therefore we shall only treat the first disjunct.

For Dominance, suppose  $\ga\vd\gb$ and $\neg(\ga\land\gb)\ev\ga$.
We have  $\neg\gb\vd\neg\ga$ and $\neg\ga\vd\neg\ga$. By  Or, we get
$\neg\gb\lor\neg\ga\vd\neg\ga$. By Supraclassicality, we have
$\neg(\ga\land\gb)\ev\neg\ga$. Applying And, we get $\neg(\ga\land\gb)\ev\bot$.

For  Conjunctiveness, suppose  $\neg(\ga\land(\ga\land\gb))\ev\ga$ and
$\neg(\gb\land(\ga\land\gb))\ev\gb$. These imply
$\neg(\ga\land\gb)\ev\ga$ and $\neg(\ga\land\gb)\ev\gb$, by Left Logical
Equivalence. Applying And, we get $\neg(\ga\land\gb)\ev\ga\land\gb$.
By reflexivity of $\vd$ and And, we have $\neg(\ga\land\gb)\ev\bot$.
By Left Logical Equivalence again, we have
$\neg(\ga\land(\ga\land\gb))\ev\bot$, and so
$\neg(\gb\land(\ga\land\gb))\ev\bot$, as desired.

For Transitivity, let $\ga\leq\gb$ and $\gb\leq\gc$. Suppose
$\neg(\ga\land\gb)\ev\bot$. By Lemma~\ref{lem:addrules}.\ref{rule:orbot}, we have
$\neg\gb\ev\bot$. Lemma~\ref{lem:addrules}.\ref{rule:botand} gives
$\neg(\gb\land\gc)\land\neg\gb\ev\bot$. By S, we have
$\neg(\gb\land\gc)\ev\gb$. So, we  have $\neg(\gb\land\gc)\ev\bot$. Using
the initial hypothesis and Or, we get $\neg\ga\lor\neg\gb\lor\neg\gc\ev\bot$. By
Lemma~\ref{lem:addrules}.\ref{rule:orbot}, we have $\neg\ga\lor\neg\gc\ev\bot$,
i.e. $\neg(\ga\land\gc)\ev\bot$.
Now, suppose  $\neg(\gb\land\gc)\ev\bot$ and $\neg(\ga\land\gc\notev\ga$.
Then, by Lemma~\ref{lem:addrules}.\ref{rule:orbot}, we have $\neg\gc\ev\bot$. Since
$\neg\ga\ev\neg\ga$, we have $\neg\ga\lor\neg\gc\ev\neg\ga$.  Therefore if
$\neg(\ga\land\gc)\ev\ga$, then And gives $\neg(\ga\land\gc)\ev\bot$.

We shall now show that the initial rational inference relation $\ev$ and the
induced one $\ev_\leq$ by the expectation ordering with ($\bf C$) are the same.

We show first that $\ev\sbs\ev_\leq$. Let $\ga\ev\gc$. We must show that
$\ga\ev_\leq\gc$. If $\gd\leq\neg\ga$, for all $\gd\in\cl$, then it clearly holds.
If not, let $\gb\equiv\neg\ga\lor\gc$ then $\vd\ga\land\gb\liff\ga\land\gc$.
So $\ga\land\gb\vd\gc$. Also,  $\ga\lor\neg\gb\equiv\ga$.
If $\ga\ev\bot$ then $\gd\leq\neg\ga$, for all $\gb\in\cl$ (using
Lemma~\ref{lem:ev-to-order}.\ref{botev}), so, by our hypothesis,
$\ga\lor\neg\gb\notev\bot$. Observe that $\ga\lor\neg\gb\ev\gc$,
and $\gc\vd\neg\ga\lor\gc\equiv\gb$. Right Weakening gives
$\ga\lor\neg\gb\ev\gb$.
So  $\gb\not\leq\neg\ga$ and therefore $\neg\ga<\gb$. Hence $\ga\ev_\leq\gc$.

For the other direction, i.e. $\ev_\leq\sbs\ev$, let $\ga\ev_\leq\gc$.
Suppose first that $\gb\leq\neg\ga$ for all $\gb\in\cl$. Therefore $\top\leq\neg\ga$.
This gives either $\neg(\neg\ga\land\bot)\ev\bot$ or
$\neg(\neg\ga\land\bot)\notev\top$. Since obviously the latter does not hold we
must have that $\ga\ev\bot$ and hence, by Right Weakening, $\ga\ev\gc$.
Now suppose that there exists $\gb\in\cl$ with $\ga\land\gb\vd\gc$ and
$\neg\ga<\gb$, i.e. $\gb\not\leq\neg\ga$. The latter implies that
$\ga\lor\neg\gb\notev\bot$ and $\ga\lor\neg\gb\ev\gb$. Observe that
$\ga\land\gb\equiv (\ga\lor\neg\gb)\land\gb$ hence, by Supraclassicality,
$(\ga\lor\neg\gb)\land\gb \ev\gc$. Applying Cut on the latter and
$\ga\lor\neg\gb\ev\gb$ we get $\ga\lor\neg\gb\ev\gc$. Applying And on
$\ga\lor\neg\gb\ev\gb$ and $\ga\lor\neg\gb\ev\ga\lor\neg\gb$ we get
$\ga\lor\neg\gb\ev\ga$. Since $\ga\lor\neg\gb\notev\bot$ we have that
$\ga\lor\neg\gb\notev\neg\ga$. Applying Rational Monotonicity on the latter and
$\ga\lor\neg\gb\ev\gc$ we get $(\ga\lor\neg\gb)\land\ga\ev\gc$, i.e. $\ga\ev\gc$.

It remains to show $\leq={\bf O}({\bf C}(\leq))$. Let $\leq'$ be ${\bf O}({\bf
C}(\leq))$ and assume $\ga\leq'\gb$. By definition of $\leq$, we have that
$\neg(\ga\land\gb)\ev\bot$ or $\neg(\ga\land\gb)\notev\ga$, where $\ev={\bf C}(\leq)$.
The former implies
$\gc\leq\ga\land\gb$, for all $\gc\in\cl$, by Lemma~\ref{lem:ev-to-order}. By
Dominance and Transitivity, we have $\ga\leq\ga\land\gb\leq\gb$, as desired. The latter
implies that
$\neg(\ga\land\gb)\lto\ga\leq\ga\land\gb$. Since  $\neg(\ga\land\gb)\lto\ga$ is
(classically) equivalent to $\ga$, we get $\ga\leq\ga\land\gb$ and so $\ga\leq\gb$,
by Dominance and Transitivity.

For the other direction, assume $\ga\leq\gb$. If $\gc\leq\ga\land\gb$, for all
$\gc\in\cl$ then $\neg(\ga\land\gb)\ev\bot$, by
Lemma~\ref{lem:ev-to-order}.\ref{botord}, and therefore $\ga\leq'\gb$, by
Definition ({\bf O}).  If not then, by Conjunctiveness on the hypothesis, we have
$\ga\leq\ga\land\gb\leq\gb$. Since  $\neg(\ga\land\gb)\lto\ga$ is
(classically) equivalent to $\ga$, we get $\neg(\ga\land\gb)\lto\ga\leq\ga\land\gb$.
So $\ga\land\gb\not<\neg(\ga\land\gb)\lto\ga$ and so, by Definition ({\bf C}) and
Lemma~\ref{lem:ev-to-order}.\ref{rest}, we have $\neg(\ga\land\gb)\notev\ga$. The
latter implies, by Definition ({\bf O}), $\ga\leq'\gb$, as desired.

\end{proof}

\begin{aproposition}
{\bf \ref{prop:complete}}\quad
A  ranked consequence operator $\ev$ based on $\vd$
is induced by a chain of sets of formulas if and only if it is induced by
the closure of this chain under arbitrary unions and intersections.
\end{aproposition}

\begin{proof}
Let $\ev$ and $\ev'$ be the ranked consequence operators induced by
$\{A_i\}_{i\in I}$ and $\{A_j\}_{j\in J}$ respectively, where the latter is the
closure of the former under arbitrary unions and intersections. Without loss of
generality we can assume that the sets belonging in $\{A_i\}_{i\in I}$ carry
the same indices in  $\{A_j\}_{j\in J}$. We must prove that $\ev$ is equal to
$\ev'$.

From left to right, suppose $\ev\gb$ and there exists $i\in I$ such that
$\ev_i\gb$. Since $i\in J$ we also have $\ev'\gb$. Suppose $\ga\ev\gb$.
If $\ga\ev_i\gb$ for some $i\in i$ then as above $\ga\ev'\gb$. If
$\ev_i\neg\ga$ for all $i$, i.e. $\neg\ga\in A_i$ for all $i$, then $\neg\ga\in
A_j$ for all
$j\in J$. For either
$A_j=\bigcup_{k}A_{i_k}$ or  $A_j=\bigcap_k A_{i_k}$, where
$i_k\in I$, and $\neg\ga\in A_{i_k}$ for all $k$.

From right to left, suppose $\ev'\gb$, then there is $j\in J$ such that
$\ev_j\gb$. Either
$A_j=\bigcup_{k}A_{i_k}$ or  $A_j=\bigcap_k A_{i_k}$, where
$i_k\in I$. In both cases there is some ${k_0}$ such that $\gb\in
A_{i_{k_0}}$. Therefore $\ev\gb$. Suppose now that $\ga\ev'\gb$.
If there exists $j\in J$ such that $\neg\ga\not\in A_j$ and $\ga\to\gb\in A_j$
then  either
$A_j=\bigcup_{k}A_{i_k}$ or  $A_j=\bigcap_k A_{i_k}$, where
$i_k\in I$. In the first case there exists ${k_0}$ such that $\ga\to\gb\in
A_{i_{k_0}}$. We also have that $\neg\ga\not\in A_{i_k}$ for all $k$. So for the
same $k_0$ we have that $\neg\ga\not\in A_{i_{k_0}}$. Therefore $\ga\ev\gb$.
In the second case we have that $\ga\to\gb\in A_{i_k}$ for all $k$ while there
exists $k_0$ such that $\neg\ga\in A_{i_{k_0}}$.
If $\neg\ga\in A_j$ for all $j\in J$, then we immediately have that $\neg\ga\in
A_i$ for all $i\in I$ and hence $\ga\ev\gb$.
\end{proof}

\begin{aproposition}{\bf \ref{prop:rank-to-rational-order}}\quad
An ordering induced by a ranked consequence operator is rational.
\end{aproposition}

\begin{proof}
Let $\langle \{B_i\}_{i\in I}, \ev \rangle$ be a ranked consequence operator. Denote
${\bf O} (\ev)$ with $\leq$, and ${\bf C} (\leq)$ with $\ev_{\leq}$.

We should verify
that
$\leq$ satisfies Supraclassicality, Transitivity, and Conjunctiveness.

For Supraclassicality, suppose $\ga\vd\gb$. We have $\vd\ga\lto\gb$, so if
$B_i\vd\ga$ than $B_i\vd\gb$, for all $i\in I$. Hence $\ga\leq\gb$.

For Transitivity, suppose $\ga\leq\gb$ and $\gb\leq\gc$. Pick an $i\in I$ such that
$B_i\vd\ga$. We have $B_i\vd\gb$, since $\ga\leq\gb$. Hence $B_i\vd\gc$, since
$\gb\leq\gc$, as desired.

For Conjunctiveness, suppose $\ga\not\leq\ga\land\gb$ and $\gb\not\leq\ga\land\gb$,
towards a contradiction. By our assumptions, there exist $B_i$ and $B_j$ with $i,j\in
J$ such that $B_i\vd\ga$ and $B_i\not\vd\ga\land\gb$ and $B_j\vd\gb$ and
$B_j\not\vd\ga\land\gb$. Now, $B_i$'s form a chain under inclusion, so either
$B_j\sbs B_i$, or $B_i\sbs B_j$. If $B_j\sbs B_i$ then $B_i\vd\gb$, a contradiction,
since $B_i\vd\ga$ and $B_i\not\vd\ga\land\gb$. Similarly, for $B_i\sbs B_j$.

We must now show that $\ga\ev\gb$ iff $\ga\ev_{\leq}\gb$. Assume $\ga\ev\gb$. We have
either $\ev_i \neg\ga$, for all $i\in I$, or there exists $i\in I$ such that
$B_i\not\vd\neg\ga$ and $B_i\vd\ga\lto\gb$. Assume the former. We have immediately
$\gc\leq\neg\ga$, for all $\gc\in\cl$. Hence $\ga\ev_\leq$, by
Lemma~\ref{lem:ev-to-order}.\ref{botord}. Assume the latter, then
$\ga\lto\gb\not\leq\neg\ga$, that is, $\neg\ga < \ga\lto\gb$. Hence $\ga\ev\gb$, by
Lemma~\ref{lem:ev-to-order}.\ref{rest}. The other direction is similar.
\end{proof}

\begin{acorollary}{\bf \ref{corol:rank-to-rational}}\quad
A ranked consequence operator is a rational inference relation.
\end{acorollary}

\begin{proof}
We shall give an alternative proof with a straightforward verification of the rules of
rational inference.
We shall show that a ranked consequence operator $\ev$ based on $\vd$ induced by
a chain of sets
$ \{B_i\}_{i\in I}$ satisfies Supraclassicality,
Left Logical Equivalence, Right Weakening, And, Cut, Cautious Monotonicity, Or
and Rational Monotonicity.

For Supraclassicality, suppose  $\ga\vd\gb$. We have either $\ev_i\neg\ga$ for
all $i\in I$ or there exists some $i\in I $ such that $\notev_i\neg\ga$.
In the first case we have immediately $\ga\ev\gb$. In the second case we have
that $B_i,\ga\vd\gb$, by our hypothesis, and therefore
$\ga\ev\gb$.

For Left Logical Equivalence, suppose that $\vd\ga\equiv\gb$ and $\ga\ev\gc$. We
have either $\ev_i\neg\ga$, i.e. $B_i\vd\neg\ga$, for
all $i\in I$ or there exists some $i\in I $ such that $\notev_i\neg\ga$ and
$B_i,\ga\vd\gc$. Since $\ga$ and $\gb$ are equivalent under
$\vd$, we have in the first case that  $B_i\vd\neg\gb$ for
all $i\in I$ and In the second case that there exists some $i\in I $ such that
$\notev_i\neg\gb$ and
$B_i,\gb\vd\gc$. In both cases we have $\gb\ev\gc$.

For Right weakening, suppose $\ga\ev\gb$ and $\vd\gb\to\gc$. If $\ev_i\neg\ga$
 for all $i\in I$ then we immediately get $\ga\ev\gc$. If there exists $i\in I$
such that $\notev_i\neg\ga$ and
$B_i,\ga\vd\gb$ then we also have that $B_i,\ga\vd\gb\to\gc$ by hypothesis.
Therefore $B_i,\ga\vd\gc$ and $\ga\ev\gc$.

For And, suppose $\ga\ev\gb$ and $\ga\ev\gc$. If $\ev_i\neg\ga$
 for all $i\in I$ then we immediately have  $\ga\ev\gb\land\gc$. If not then
there exists $i,k\in I$ such that $\notev_i\neg\ga$, $\notev_k\neg\ga$,
$B_i,\ga\vd\gb$, and $B_k,\ga\vd\gc$.
Since $\leq$ is linear let $i\leq k$. Then $B_i\sbs B_k$ and therefore
$B_k,\ga\vd\gb$. So $B_k,\ga\vd\gb\land\gc$ and $\ga\ev\gb\land\gc$.

For Cut, suppose $\ga\ev\gb$ and $\ga\land\gb\ev\gc$. If $\ev_i\neg\ga$ for all
$i\in I$ then we immediately have $\ga\ev\gc$. If not, then there exists  $i\in I$
such that $\notev_i\neg\ga$ and $B_i,\ga\vd\gb$. If
$\ev_j\neg(\ga\land\gb)$( $\equiv\neg\ga\lor\neg\gb$) for all $j\in I$, then for
$i$,
$B_i\vd\neg\ga\lor\neg\gb$, i.e.
$B_i,\ga\vd\neg\gb$. Combining with our hypothesis we get
$B_i\vd\neg\ga$, i.e. $\ev_i\neg\ga$, a contradiction.
Therefore there exists $k\in I$ such that $\notev_k \neg\ga\lor\neg\gb$ and
$B_k,\ga\land\gb\vd\gc$. There are two cases: either $k\leq i$
or $i\leq k$. In the first case we have
$B_i,\ga\land\gb\vd\gc$ as well, so by (regular) cut on $\vd$
and our hypothesis we get $B_i,\ga\vd\gc$. Therefore
$\ga\ev\gc$. In the second case, observe that $B_k,\ga\vd\gb$ so as above
$B_k,\ga\vd\gc$. Since
$B_k\not\vd \neg\ga\lor\neg\gb$ then
$B_k\not\vd \neg\ga$ so again $\ga\ev\gc$.

For Cautious Monotony, suppose $\ga\ev\gb$ and $\ga\ev\gc$. If $\ev_i \neg\ga$
for all $i\in I$ then we also have $\ev_i\neg\ga\lor\neg\gb$ hence
$\ev_i\neg(\ga\land\gb)$ for all $i\in I$. Therefore $\ga\land\gb\ev\gc$.
If not, there exists $i,k\in I$ such that $\notev_i\neg\ga$,
$B_i,\ga \vd\gb$, $\notev_j\neg\ga$ and $B_j, \ga\vd\gc$. Let $l=\max(i,k)$ then
$B_k\not\vd\neg\ga$ and both $B_l,\ga\vd\gb$ and $B_l,\ga\vd\gc$. From the
latter we get $B_l,\ga\land\gb\vd\gc$. Suppose that
$B_k\vd\neg\ga\lor\neg\gb$ then $B_l,\ga\vd\neg\gb$. Combining it with above we
get $B_l\vd\neg\ga$ which is a contradiction to our hypothesis.

For Or, suppose that $\ga\ev\gc$ and $\gb\ev\gc$. If both $\ev_i\neg\ga$  and
$\ev_i\vd\neg\gb$ for all $i\in I$ then we also have $\ev_i\neg\ga\land\neg\gb$
for all $i$ and therefore $\ga\lor\gb\ev\gc$.
If this is true for only one of them, say $\ev_i\neg\ga$ for all $i\in I$ but
there exists $k\in I$ such that $\notev_k\neg\gb$ and $B_k,\gb\vd\gc$, then we also
have
$\notev_k\neg\gb\land\neg\ga$ and $B_k$ which implies $B_k,\ga\vd\gc$. So
$B_k,\ga\lor\gb\vd\gc$ and therefore $\ga\lor\gb\ev\gc$.
If neither of them holds then there exist $i,k\in I$ such that $\notev_i\neg\ga$,
$B_i,\ga\vd\gc$, $\notev_k\neg\gb$ and $B_k,\gb\vd\gc$. Let $l=\max(i,k)$ then we
have
$\notev_l\neg\ga\lor\neg\gb$, $B_l,\ga\vd\gc$ and $B_l,\gb\vd\gc$. So
$B_l,\ga\lor\gb\vd\gc$ and therefore $\ga\lor\gb\ev\gc$.

For Rational Monotonicity, suppose that $\ga\notev\neg\gb$ and $\ga\ev\gc$. We
can't have $\ev_i\neg\ga$ for all $i\in I$ because we have that
$\ga\notev\neg\gb$. So there exists $i\in I$ such that $\notev_i\neg\ga$ and
$B_i,\ga\vd\gc$. We have by monotonicity of $\vd$ that $B_i,\ga\land\gb\vd\gc$.
Now observe that $\ga\notev\neg\gb$ implies that if $B_i,\ga\vd\neg\gb$ then
$\ev_i\neg\ga$. However since $\notev_i\neg\ga$ we have that
$B_i,\ga\not\vd\neg\gb$ so $B_i\not\vd\neg\ga\lor\neg\gb$. Therefore
$\ga\land\gb\ev\gc$ by definition.

\end{proof}

\begin{atheorem}{\bf \ref{thm:rational-order-to-rank}}\quad
A  rational inference relation is a ranked consequence operator.
\end{atheorem}

\begin{proof}
Denote the comparative rational inference relation by $\ev_\leq$. We shall define
a chain of sets $\{A_i\}_{i\in I}$ which generates a ranked consequence
operator $\ev$ equal to $\ev_\leq$. Let $\sim$ be the equivalence relation
induced by $\leq$ (an expectation ordering is clearly a preorder). The
equivalence classes will be denoted by $\hat{\ga}$ (where $\ga\in\hat{\ga}$). It
is also clear that the set of equivalence classes is linearly ordered.
Now, for each $\ga\in\cl$,  let
$$A_{\hat{\ga}}=\{\gb : \ga\leq\gb\}.$$
Note here that, by Dominance, the sets $A_{\hat{\ga}}$ are closed under
consequence. Moreover, we have $A_{\hat{\ga}}\sbs A_{\hat{\gb}}$ iff $\gb\leq\ga$.
Now, generate a ranked consequence operator $\ev$ as in Definition~\ref{def:rankvd}.

We must show that $\ev$ and $\ev_\leq$ are identical.

Let $\ga\ev_\leq\gb$, i.e. either $\gb\leq\neg\ga$ for all $\gb\in\cl$, or there
exists $\gd\in\cl$ such that $\neg\ga<\gd$ and $\gd\land\ga\vd\gc$. In the first
case $\neg\ga\in A_{\hat{\gb}}$ for all $\gb\in\cl$. So $\ev_{\hat{\gb}} \neg\ga$
for all $\gb\in\cl$ and therefore $\ga\ev\gc$.
in the second case consider $A_{\hat{\gd}}$. We have $\gd\in A_{\hat{\gd}}$ so
$A_{\hat{\gd}},\ga\vd\gc$. Suppose that $A_{\hat{\gd}}\vd\neg\ga$ then by
compactness there exists $\gd'\in A_{\hat{\gd}}$ such that $\gd'\vd\neg\ga$. By
Dominance we have $\gd'\leq\neg\ga$ and by definition of $A_{\hat{\gd}}$,
$\gd\leq\gd'$, i.e. $\gd\leq\neg\ga$, a contradiction.

Now let $\ga\ev\gc$. If $\ev_{\hat{\gb}}\neg\ga$ for all $\gb\in\cl$ then
$\gb\leq\neg\ga$ for  all $\gb\in\cl$ and we are done. If not, then there exists
${\hat{\gd}}$ such that $A_{\hat{\gd}}\not\vd\neg\ga$ and $A_{\hat{\gd}},\ga\vd\gc$.
Since
$A_{\hat{\gd}}\not\vd\ga$ we must have $\neg\ga<\gd$. By compactness there exists
$\gd'\in A_{\hat{\gd}}$ such that $\gd',\ga\vd\gc$, i.e. $\gd'\land\ga\vd\gc$ and
$\neg\ga<\gd\leq\gd'$. Therefore $\ga\ev\gc$.
\end{proof}

\end{document}